\documentclass[a4paper]{spie}

\usepackage{amsmath,amsfonts,amssymb}
\usepackage{graphicx}
\usepackage{siunitx}
\usepackage[version=4]{mhchem}
\usepackage[colorlinks=true, allcolors=blue]{hyperref}
\usepackage[capitalize]{cleveref}

\usepackage{sidecap}

\title{Additively manufactured ceramic platforms for miniaturized optical frequency references and robust atomic sensor systems}

\author[a,*]{M. Christ}
\author[a,b]{S. Neinert}
\author[a]{C. Zimmermann}
\author[a,b]{K. D\"oringshoff}
\author[a,b]{M. Krutzik}
\affil[a]{Ferdinand-Braun-Institut (FBH), Gustav-Kirchhoff-Stra\ss{}e 4, 12489 Berlin, Germany}
\affil[b]{Humboldt-Universit\"at zu Berlin, Institute of Physics \& CSMB, Newtonstra\ss{}e 15, 12489 Berlin, Germany }

\authorinfo{*Corresponding author: Marc Christ, marc.christ@fbh-berlin.de, www.fbh-berlin.de}

\begin{document} 
\maketitle

\begin{abstract}
Transferring atomic quantum sensors from laboratory-scale physics packages to compact, alignment-free sensor heads suitable for field and space deployment requires a structural platform that combines mechanical, thermal, and optical functionality within a single, manufacturable component. We present such a platform based on additively manufactured technical ceramics: lithography-based ceramic manufacturing is used to fabricate monolithic micro-optical benches with application-specific mounting geometries, while high-precision active alignment and adhesive bonding realize long-term stable optical assemblies through micro-integration. As a first system-level demonstrator, we present a micro-integrated optical frequency reference based on Doppler-free spectroscopy of the rubidium \ce{D2} line, integrated within a \SI{6}{\milli\liter} \ce{Al2O3} module. A laser stabilized to the reference reaches a fractional frequency instability of $3\times10^{-12}$ at \SI{1}{\second}. We extend the platform to a compact optically pumped magnetometer, exploiting the magnetic cleanliness and tailored thermal properties of high-purity ceramics for a \SI{7}{\milli\liter} sensor head with picotesla sensitivity aimed at space-borne biomagnetometry, and to a micro-integrated optical dipole trap with active MEMS-mirror beam steering for cold-atom manipulation. Together, these demonstrators show that ceramic additive manufacturing and optical micro-integration form a versatile, robust platform for compact atomic sensor systems in field and space applications.
\end{abstract}


\section{Introduction}
\label{sec:introduction}

Atom-based quantum sensors exploit well-defined atomic states and transitions to measure physical quantities such as magnetic and electric fields, gravity, and inertial forces with high sensitivity and accuracy, or to serve as frequency references in precision timing systems \cite{Kitching2011,Bongs2019,Degen2017}. Their applications span precision metrology and navigation, resource exploration, and fundamental physics. In space, such systems are considered for next-generation timing and navigation infrastructure \cite{Schuldt2021}, with optical frequency references already demonstrated in sounding-rocket experiments \cite{Dinkelaker2017,Doeringshoff2019} and optical atomic clocks currently under development for in-orbit demonstration \cite{Kuschewski2024}. Cold-atom sensors are further investigated for future satellite-based Earth observation missions \cite{Leveque2022}. Atomic magnetometers offer access to weak biomagnetic signals, such as magnetic fields generated by brain or muscle activity, providing a route towards non-invasive medical diagnostics without cryogenic sensor operation \cite{Boto2018,Tierney2019,Aslam2023}. As discussed in \cref{sec:opm}, this capability is of particular interest for astronaut health monitoring on crewed space missions. These examples illustrate the broad potential of atomic quantum systems beyond specialized laboratory environments.

Realizing this potential in practical field and space applications, however, requires a substantial reduction in size, weight, power consumption, and system complexity. The sensor's core subsystem, commonly referred to as the physics package, typically combines an optical system for atomic state preparation, manipulation, and detection with the atomic ensemble, contained in a vapor cell or vacuum system. Additional subsystems tailored to the specific sensing modality, such as field generation or thermal control, are integrated alongside these core components. Because sensor performance is highly sensitive to optical alignment, the physics package must remain mechanically and thermally stable under environmental loads. Conventional laboratory implementations rely on adjustable optomechanics and comparatively long free-space beam paths. While highly flexible during development, such assemblies require expert alignment and maintenance and must instead be transferred into compact, alignment-free subsystems and sensor heads suitable for mobile deployment.

Micro-integration provides an established route towards compact and mechanically stable optical assemblies. Electro-optical components with characteristic dimensions of only a few millimeters can be actively aligned and bonded onto a micro-optical bench (MIOB), resulting in compact systems that no longer require adjustable optomechanics after assembly \cite{Schiemangk2015,Kuerbis2020,Christ2024a}. Related integration technologies have been successfully employed in drop-tower \cite{Muntinga2013} and sounding-rocket experiments \cite{Dinkelaker2017,Becker2018a,Doeringshoff2019} and are planned for further space missions \cite{Frye2021} aboard the ISS. For application in such demanding environments, the optical integration process must be reproducible and robust. Adhesive bonding technologies used for micro-integration have therefore been subjected to dedicated environmental qualification campaigns \cite{Christ2023}.

The MIOB is not only a mechanical carrier but a central, functional part of the optical system, integrating the various mechanical, optical, thermal, and electrical features of the physics package. To keep component placement and optical alignment stable under mechanical and thermal loads, its material must combine high stiffness, low mass, and a low coefficient of thermal expansion (CTE). Technical ceramics are particularly attractive for this purpose because they combine high stiffness and temperature stability with low density, electrical insulation, vacuum compatibility, magnetic cleanliness, and a broad range of available thermal properties. These characteristics are relevant for robust optical and electronic assemblies and make technical ceramics an interesting material class for compact field- and space-oriented sensor systems. Detailed material properties and a comparison with other materials commonly used in compact quantum systems are given in Ref.~\citenum{Christ2024}.

Additive manufacturing (AM) complements these material properties with a high degree of geometric freedom, overcoming key limitations of traditional ceramic manufacturing such as restricted design complexity, high tooling costs, and long production times. Because it avoids per-design tooling, AM is particularly cost-effective for complex geometries across a wide range of production volumes, from single, highly customized components such as space instrumentation to larger production series. Complex three-dimensional mounting structures, internal cavities, lightweight structures, and application-specific interfaces can be incorporated directly and efficiently into a monolithic ceramic part. This geometric freedom further enables topology optimization approaches, supporting lightweight designs, tailored thermal properties, and material savings. In previous work, we introduced lithography-based ceramic manufacturing (LCM) for compact quantum technologies and demonstrated its use for a micro-integrated optical frequency reference \cite{Christ2024}. Here, we present the ceramic AM and micro-integration approach as a broader technology platform for compact optical and atomic sensor systems. After introducing the manufacturing and integration concept, we discuss the optical frequency reference CerAMRef as an integrated demonstrator and subsequently extend the approach to compact optically pumped magnetometers and optical systems with integrated active beam-steering elements for cold-atom manipulation.

\section{Ceramic additive manufacturing and micro-integration platform}
\label{sec:ceramic_am}

\subsection{Technical ceramics and lithography-based ceramic manufacturing}
\label{subsec:lcm}

Additive manufacturing has increasingly been investigated for components of compact quantum systems, including metallic vacuum parts, magnetic coils and shields, and polymer structures for mounting optical components \cite{Norrgard2018a,Saint2018,Vovrosh2017,Madkhaly2022}. Technical ceramics complement these printable material classes with properties especially advantageous for precise optical and atomic sensor assemblies deployed on mobile platforms or in space. LCM enables printing of various oxide and nitride ceramics, including alumina (\ce{Al2O3}), zirconia (\ce{ZrO2}), aluminum nitride (\ce{AlN}), and silicon nitride (\ce{SiN}), with mechanical and thermal properties comparable to traditionally manufactured ceramics: \ce{Al2O3} combines high stiffness with moderate density, electrical insulation, temperature stability, and vacuum compatibility, while \ce{ZrO2} and \ce{AlN} offer, respectively, thermal isolation and high thermal conductivity for applications requiring substantially different thermal behavior. A detailed comparison of relevant properties for additively manufactured materials is provided in Ref.~\citenum{Christ2024}. Ceramic components for the presented physics packages in this work are fabricated by LCM using a CeraFab S65 system (Lithoz GmbH): a vat-photopolymerization process in which a suspension of ceramic particles in a photosensitive binder is selectively polymerized layer by layer using a digital micromirror device \cite{Schwentenwein2014,Schwentenwein2015}. The printer provides a lateral discretization of \SI{40}{\micro\meter} over a build area of \qtyproduct{102 x 64}{\milli\meter} and supports layer thicknesses between \SI{10}{\micro\meter} and \SI{100}{\micro\meter}; a layer thickness of \SI{25}{\micro\meter} is typically used for the MIOBs discussed here \cite{Christ2024}.

\begin{figure}[tbp]
    \centering
    \includegraphics[width=\linewidth]{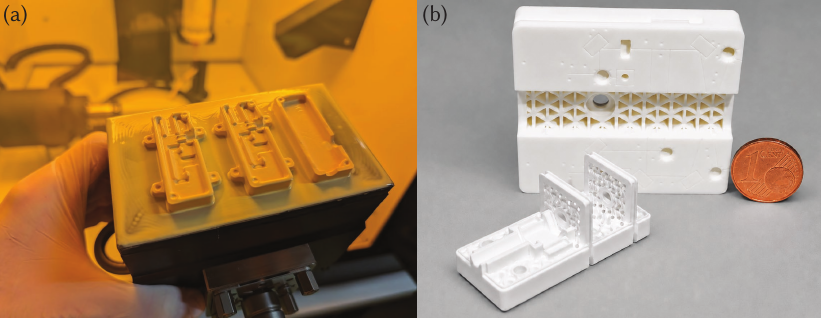}
    \caption{\label{fig:lcm_platform} Ceramic additive manufacturing for compact optical and atomic systems.
    (a) Exemplary build platform with printed MIOBs after completion of an LCM print job.
    (b) Two fully processed MIOBs, fabricated from \ce{Al2O3} and \ce{ZrO2}, respectively, featuring gyroid and isogrid structures for weight optimization as well as the discussed geometries for mounting and alignment of (electro-)optical components and vapor cells, shown with a one-euro-cent coin for scale.}
\end{figure}

After printing, uncured slurry is cleaned off the parts and the resulting green parts, which consist of ceramic particles held together by the cured polymer binder, are thermally processed: the binder is removed during debinding before the ceramic is densified by sintering, with the CAD geometry scaled prior to printing to compensate for the resulting material-specific shrinkage. For oxide ceramics, including \ce{Al2O3} and \ce{ZrO2}, the complete process chain from printing and cleaning through debinding and sintering is operated in-house, enabling rapid iteration from component design, fabrication, and geometric inspection to system assembly. \cref{fig:lcm_platform}(a) shows printed parts on the build platform, while \cref{fig:lcm_platform}(b) shows representative MIOBs after complete processing. In addition to \ce{Al2O3} structures, \ce{ZrO2} has been introduced for applications in which thermal isolation is an integral part of the mechanical design. Examples include compact vapor-cell ovens and thermally insulating MIOBs. In this way, the ceramic material is selected to match the specific requirements of each application.

\subsection{Design for optical micro-integration}
\label{subsec:microintegration}

A major advantage of ceramic AM for optical systems is the ability to combine the mechanical base structure and component-specific mounting geometries within a monolithic part. Instead of using a planar substrate with separately manufactured mounts, features such as V-grooves, pockets, alignment edges, recesses, sidewalls, cable or fiber guides, and interfaces for sensors can be incorporated directly into the optical bench, allowing irregular components such as glass-blown atomic vapor cells to be accommodated by geometries adapted to their actual dimensions. These printed features provide mounting surfaces and, in some cases, mechanical pre-alignment for the optical components, together with reference geometries that define the beam path layout. The final component positions are subsequently established during micro-integration on the respective mounting surface. Thus the dimensional precision of LCM itself does not need to provide a final optical alignment accuracy. For the previously characterized \ce{Al2O3} optical benches, dimensional deviations of relevant features remained on the order of several tens of micrometers \cite{Christ2024}, sufficient to position millimeter-scale optical components well within the range of the subsequent active alignment process.

Micro-integration is performed using precision positioning systems that manipulate up to four optical and/or electro-optical components simultaneously in all six degrees of freedom, with actuator step sizes down to \SI{1}{\nano\meter} and \SI{1}{\micro\radian} \cite{Christ2023,Christ2024a}. The optical subsystem is operated during assembly and the relevant optical signal is used directly as the alignment criterion. Aligned components are subsequently fixed to the ceramic substrate by thermal- or UV-curing adhesive bonding, qualified for micro-integrated optical systems under environmental and vacuum requirements \cite{Christ2023}. The bonding permanently transfers the optical alignment into a mechanically stable assembly without adjustable mounts. This combination of application-specific ceramic geometries, active optical alignment, and qualified bonding has already been demonstrated for direct cold-atom manipulation in a micro-integrated crossed-beam optical dipole trap on a ceramic MIOB, showing long-term alignment stability and robustness under operational conditions \cite{Christ2024a}.

\section{Micro-integrated optical frequency reference}
\label{sec:ceramref}

As a first system-level demonstrator combining ceramic AM and optical micro-integration, we developed a micro-integrated optical frequency reference, referred to as CerAMRef, based on Doppler-free spectroscopy of the rubidium \ce{D2} line at \SI{780}{\nano\meter} \cite{Christ2024}. Such references are required for stabilizing laser sources used in atomic-physics experiments and quantum sensors and are conventionally realized using free-space spectroscopy setups occupying substantially larger volumes.

\begin{figure}[tbp]
    \centering
    \includegraphics[width=\linewidth]{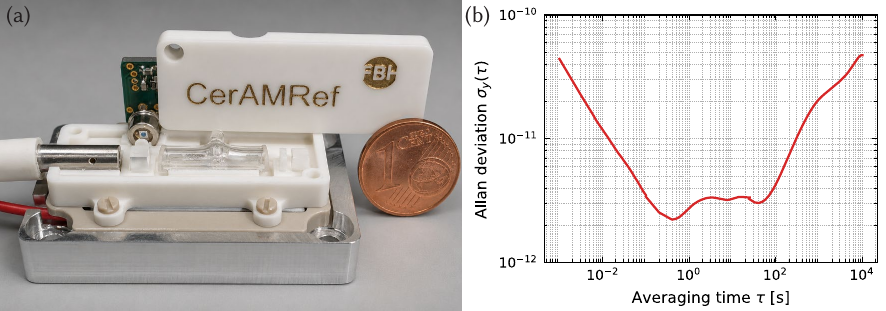}
    \caption{\label{fig:ceramref} CerAMRef optical frequency reference.
    (a) Assembled \SI{6}{\milli\liter} module with an additively manufactured \ce{Al2O3} bench and lid. The fiber-coupled optical system, rubidium vapor cell, micro-optical components, and photodiode are integrated directly on the printed ceramic bench. A one-euro-cent coin is shown for scale.
    (b) Overlapping Allan deviation of the beat-note frequency between a laser stabilized to the CerAMRef and an independent rubidium reference. Data replotted from the measurements reported in Ref.~\citenum{Christ2024}.}
\end{figure}

The CerAMRef implements a Doppler-free retro-reflection spectroscopy geometry in a compact, fiber-coupled layout with a volume of approximately \SI{6}{\milli\liter}, shown in \cref{fig:ceramref}(a). Light enters the module through a fiber collimator with a polarization-maintaining fiber and passes a polarizing beam splitter (PBS) before propagating through a rubidium vapor cell. A quarter-wave plate and dielectric mirror generate a counter-propagating second pass through the cell. After returning through the quarter-wave plate, the orthogonally polarized light is separated at the polarizing beam splitter and focused onto a photodiode. The resulting counter-propagating geometry enables Doppler-free spectroscopy while maintaining a short and compact optical layout. Further details of the spectroscopy configuration are given in Ref.~\citenum{Christ2024}.

The optical system is integrated on a monolithic LCM-fabricated \ce{Al2O3} MIOB. Custom mounting geometries are incorporated directly into the ceramic substrate for the fiber collimator, micro-optics, photodiode, temperature sensor, and the glass-blown rubidium cell. In particular, the geometry surrounding the vapor cell illustrates an advantage of additive manufacturing: the mounting structure can be adapted to the non-ideal geometry of an individual glass component while simultaneously defining its position relative to the optical axis. Most free-space optical components have characteristic edge lengths of approximately \SI{2.5}{\milli\meter}. They are actively aligned while monitoring the optical signal and subsequently bonded to the ceramic substrate using the qualified micro-integration approach described in Ref.~\citenum{Christ2023}. After bonding, the system forms an alignment-free optical assembly without adjustable optomechanics.

\cref{fig:ceramref}(a) shows the assembled module with its ceramic lid removed. The compact arrangement of the vapor cell, micro-optics, detector, and fiber interface illustrates the high degree of integration enabled by the printed three-dimensional substrate. A Pt100 temperature sensor is incorporated into the MIOB and the overall module is temperature stabilized using a Peltier element. A compact PCB provides the electrical interface and photodiode amplification. The ceramic housing has outer dimensions of approximately
\SI{14.8}{\milli\meter}$\times$\SI{40}{\milli\meter}$\times$\SI{9.5}{\milli\meter}.
The complete integrated spectroscopy module has a mass of \SI{15}{\gram}, including the optical components. An optical path length of approximately \SI{53}{\milli\meter} is realized inside the module. The fiber-coupled architecture allows the frequency reference to be integrated as an independent subsystem into higher-level systems.

The performance of the CerAMRef was evaluated by stabilizing a narrow-linewidth external-cavity diode laser (ECDL) to the $^{87}$Rb \ce{D2} $F=2 \rightarrow F'=2/3$ crossover transition of the Doppler-free rubidium spectrum using frequency-modulation spectroscopy (FMS), which provides a suitable dispersive signal for laser stabilization. The employed micro-integrated ECDL is based on a compact laser technology platform developed for robust field and space-oriented optical systems \cite{Kuerbis2020}. The frequency instability of the stabilized system was determined from a beat-note measurement against an independent rubidium reference based on modulation-transfer spectroscopy. The complete experimental configuration, including the optical-power stabilization, modulation electronics, and frequency evaluation procedure, is described in detail in Ref.~\citenum{Christ2024}; only the resulting frequency-reference performance is summarized here.

\cref{fig:ceramref}(b) shows the overlapping Allan deviation of the beat-note frequency. The measured fractional frequency instability reaches
$3\times10^{-12}$ at an averaging time of \SI{1}{\second} and approximately
$4\times10^{-12}$ at \SI{100}{\second}. The instability remains below
$1\times10^{-11}$ for averaging times up to approximately \SI{300}{\second} and below
$1\times10^{-10}$ up to \SI{1e4}{\second} \cite{Christ2024}.

These results are typical for a rubidium \ce{D2}-line FMS reference, yet achieved within a \SI{6}{\milli\liter} module---a volume reduction of multiple orders of magnitude compared to conventional free-space implementations. This highlights how the ceramic AM platform facilitates the integration of highly compact, mechanically stable, alignment-free systems, while retaining performance sufficient for laser-frequency stabilization in cold-atom experiments and related atomic quantum sensors. The CerAMRef therefore demonstrates the central concept of the ceramic integration platform: the additively manufactured component serves as a highly stable functionalized substrate that provides the mounting and reference geometries for a complete optical subsystem, with the final optical alignment established through precision micro-integration.

\section{Compact optically pumped magnetometer}\label{sec:opm}
Optically Pumped Magnetometers (OPMs) exploit the quantum mechanical interaction between atomic spins and optical fields to achieve highly sensitive magnetic field measurements \cite{Budker2007}. In a typical alkali vapor OPM setup, resonant circularly polarized laser light is transmitted through an atomic vapor cell, transferring angular momentum from photons to valence electrons. This optical pumping process drives the atomic ensemble into a coherent, spin-polarized ground state. In the presence of external magnetic fields, the macroscopically polarized atomic spins undergo Larmor precession around the magnetic field vector at a frequency proportional to the local field strength.
This spin precession modulates the optical absorption and dispersion properties of the atomic medium at the Larmor frequency. By detecting the transmitted optical power or polarization components, minute magnetic field changes can be extracted with sub-\unit{\pico\tesla} precision. Crucially, this high-resolution measurement is achieved without requiring the cryogenic cooling infrastructure necessary for Superconducting Quantum Interference Devices (SQUIDs), making OPMs inherently more suitable for mobile applications and remote deployment \cite{Tiporlini2013}.

For OPM operation, the magnetic cleanliness of high-purity ceramics is of particular relevance: their near-zero magnetic footprint, resulting from low magnetic susceptibility, avoids local structural magnetic bias and stray field contamination that would otherwise interfere with high-sensitivity measurements. A tailored thermal design, combining application-optimized geometries with a suitably selected ceramic material, further enables stable operation at reduced power consumption, supporting compact, skin-safe sensor head concepts for wearable or body-mounted deployment.

\begin{SCfigure}[1][tb]
    \centering
    \includegraphics[width=0.7\linewidth]{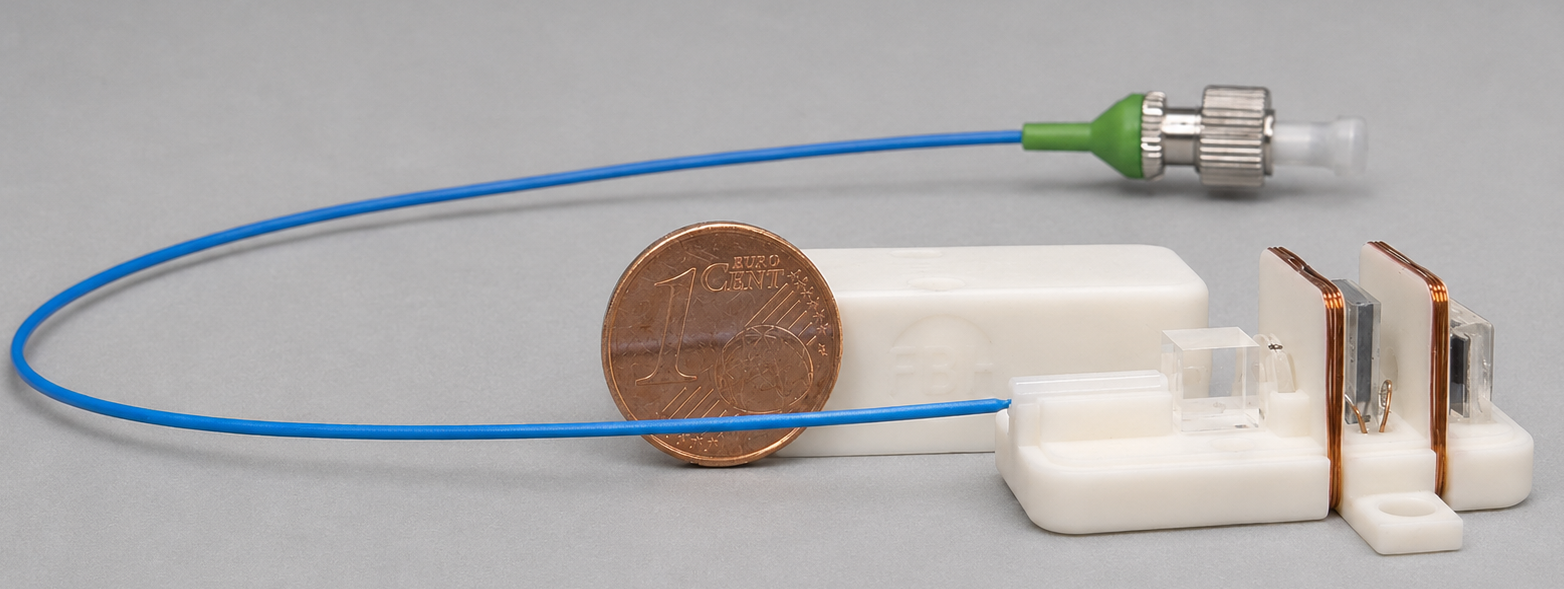}
    \caption{\label{fig:OPM} Miniaturized $M_x$-type OPM sensor head, occupying a volume of approximately \SI{7}{\milli\liter}. The core of the demonstrator unit is a monolithic MIOB additively manufactured from alumina (\ce{Al2O3}).
    }
\end{SCfigure}

The physics package of our $M_x$-type OPM in development, depicted in \cref{fig:OPM}, occupies a volume of approximately \SI{7}{\milli\liter} with a total mass of \SI{16}{\gram} and is integrated onto a ceramic MIOB. Fiber-coupled light delivery keeps the light source separate from the sensing unit, preserving magnetic cleanliness. After collimation and polarization cleaning, the circularly polarized light is transmitted through a heated cesium vapor cell. In ongoing development, the cell oven is thermally isolated using additively manufactured \ce{ZrO2} structures, reducing the stand-off distance required between the optical sensing volume and the sample under investigation compared to conventional thermal insulation approaches; the next-generation \ce{ZrO2} bench for this demonstrator is shown in \cref{fig:lcm_platform}(b). The cell is sandwiched between a pair of radio-frequency Helmholtz coils that coherently drive the spin resonance. Finally, the transmitted light is detected on a photodiode.

The primary application driving our OPM developments is non-invasive space-borne biomagnetometry. Extended exposure to microgravity during crewed space missions causes degeneration of the neuromuscular apparatus. Tracking muscle atrophy and evaluating countermeasure exercise routines inside spacecraft habitats requires continuous contactless diagnostic tools. Resolving such biomagnetic fields via magnetomyography (MMG) demands magnetic field probes capable of sub-\unit{\pico\tesla} sensitivities across measurement bandwidths up to the \unit{\kilo\hertz} range \cite{Semeia22}. First measurements with our OPM demonstrators show that they are capable of detecting magnetic fields with a sensitivity of \SI{1.2}{\pico\tesla\per\sqrt{\hertz}} at a bandwidth of \SI{260}{\hertz}.

Beyond space health monitoring and diagnostics, miniaturized magnetometers have strong potential for broader space exploration and terrestrial sensing missions. In GNSS-denied environments, such as terrestrial settings like urban canyons and underground infrastructure, or potentially on lunar or Martian surfaces, Magnetic-Aided Inertial Navigation Systems (MAINS) utilize real-time magnetic field measurements to match local crustal magnetic anomaly maps, constraining accelerometer drift in autonomous landers, rovers, and ground vehicles \cite{Huang24,Ingleby24}.

\section{Integration of active optical elements for atom trapping}\label{sec:active_optics}
\begin{SCfigure}[1][tb]
    \centering
    \includegraphics[width=0.55\linewidth]{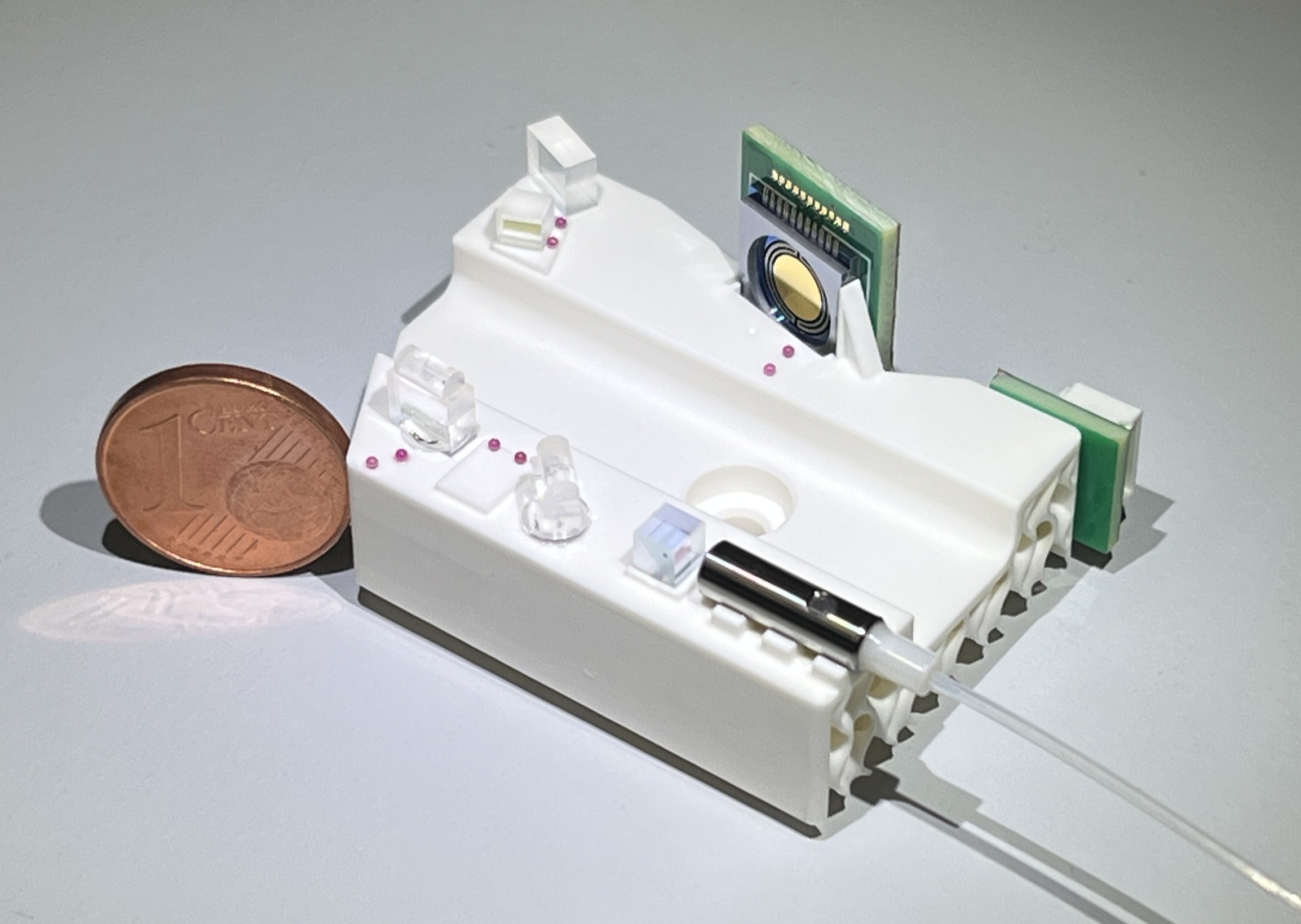}
    \caption{\label{fig:MEMSodt} Miniaturized active optical dipole trap system (volume: \SI{21}{\milli\liter}, mass: \SI{19}{\gram}) employing a 2D MEMS mirror for dynamic beam steering. The MIOB integrates a mass-reducing gyroid lattice structure alongside customized alignment and mounting features for the optical components.}
\end{SCfigure}

Cold-atom-based sensor concepts benefit from dynamic control of the optical trapping potential, e.g., by rapidly scanning a focused beam to create a time-averaged, or ``painted,'' potential, which provides trap geometries and evaporation efficiencies not accessible with static optical dipole traps \cite{Henderson_2009}. Classical beam-steering approaches for such applications, such as acousto-optic deflectors \cite{Barredo2016} or digital micromirror devices \cite{Stuart2017}, are effective but comparatively bulky and power-hungry, limiting their suitability for miniaturized, integrated optical systems. Building on our previously demonstrated static crossed-beam optical dipole trap on a ceramic MIOB \cite{Christ2024a}, we extend the platform to active optical elements by integrating a compact piezoelectric MEMS mirror for dynamic beam steering onto the ceramic bench, see \cref{fig:MEMSodt}.

The optical setup is designed around an additively manufactured \ce{Al2O3} MIOB with a compact footprint of \qtyproduct{42 x 31.5}{\milli\meter}, occupying a volume of \SI{21}{\milli\liter} with a mass of \SI{19}{\gram}. It features multiple alignment markers and mounting features for the optical components. The additive manufacturing process enables the integration of a gyroid lattice structure, reducing the substrate's mass while ensuring high mechanical rigidity and minimizing excess material. Furthermore, rapid prototyping enabled the iterative optimization of the mechanical alignment and mounting features, which are tailored to the specifications of the MEMS mirror used. The mirror is adhesively bonded onto the ceramic bench and connected to a PCB, which facilitates wire-bonded electrical connections for its operation.

A pigtailed fiber collimator delivers polarized laser light onto the bench depicted in \cref{fig:MEMSodt}. The light is further conditioned by a PBS and reflected onto a gold-coated MEMS mirror manufactured by Fraunhofer ISIT. The actively steered laser beam is subsequently guided through a focusing lens, which translates the steering angle into a lateral beam displacement by up to \qty{\pm 1}{\milli\meter} for this design. As a result, the beam focus ($w_\mathrm{0} \approx \qty{30}{\micro\meter}$) creating the dipole trap behind the optical system is dynamically displaced in its focal plane. The 2D MEMS mirror enables either one-dimensional painting along the main axes of the focal plane or combined two-dimensional painting following a Lissajous pattern, depending on driving frequencies, which are in the range of about \qtyrange{3}{9}{\kilo\hertz}.

The system is designed to operate at optical powers up to \qty{5}{\watt} continuous wave (cw) at \qty{1064}{\nano\meter} in a cold-atom experiment, where it is being evaluated for improved loading of \ce{^{87}Rb} atoms from a magneto-optical trap into the optical dipole trap. Further applications of the technology include compact ultra-cold-atom systems, creating Bose-Einstein condensates (BEC), which could enable long-lived quantum memories in space \cite{DaRos2023}.

\section{Conclusion and outlook}
\label{sec:conclusion}

Additively manufactured technical ceramics provide a versatile structural platform for the integration of compact optical and atomic quantum systems. Their combination of mechanical stiffness, temperature stability, electrical insulation, and vacuum-compatible material properties is complemented by the geometric freedom of additive manufacturing. Using LCM, application-specific mounting structures, lightweight geometries, thermal functionality, and interfaces for optical and electronic components can be incorporated directly into a monolithic MIOB, which can be manufactured effectively and scalably. In combination with active micro-integration and qualified adhesive bonding, the dimensional tolerances of the ceramic printing process are compatible with the alignment requirements of precision optical systems.

The CerAMRef demonstrates this approach at the subsystem level. Doppler-free rubidium spectroscopy is realized within an approximately \SI{6}{\milli\liter}, \SI{15}{\gram} fiber-coupled module based on a printed \ce{Al2O3} optical bench. A laser stabilized to the reference achieves a fractional frequency instability of
$3\times10^{-12}$ at \SI{1}{\second}, demonstrating that substantial miniaturization can be achieved while retaining performance suitable for cold-atom and metrology applications.

The ceramic platform is being extended both in terms of materials and integrated functionality. In addition to \ce{Al2O3}, \ce{ZrO2} is now processed for thermally isolating structures such as compact vapor-cell ovens and micro-optical benches, illustrated by the heated cesium cell of the compact optically pumped magnetometer presented in \cref{sec:opm}, where it is intended to reduce the stand-off distance to the sample. The complete printing, debinding, and sintering process chain for oxide ceramics is available in-house. Selective metallization is under development to introduce electrical interconnects and locally integrated functional elements. The compact OPM, aimed at space-borne biomagnetometry, and the active optical systems for cold-atom manipulation presented in \cref{sec:opm,sec:active_optics} further illustrate how the same manufacturing and integration approach can be transferred from a spectroscopy reference to different atomic-sensor architectures.

Future developments will increasingly combine mechanical, optical, thermal, and electrical functionality within the ceramic substrate. Such multifunctional components can reduce part count and assembly complexity while providing stable and reproducible interfaces between optical components, atomic media, and higher-level sensor systems. In parallel, the platform is also being applied towards compact quantum memory modules based on a heated atomic vapor cell, extending the micro-integration approach demonstrated for the CerAMRef and the OPM. This makes additive manufacturing of technical ceramics a promising enabling technology for the further transition of atomic quantum devices from laboratory experiments towards compact systems for field and space applications.

\acknowledgments 
This work is supported by the German Space Agency (DLR) with funds provided by the Federal Ministry of Research, Technology and Space (BMFTR) under grant numbers DLR 50WM2070, 50RK2580A, 50WM2268, and 50SI2655B.

\bibliography{bibliography_main} 

\bibliographystyle{spiebib} 

\end{document}